\documentclass[aps,reprint,prb,twocolumn,showpacs,preprintnumbers]{revtex4}
\usepackage{graphics}
\usepackage{graphicx}% Include figure files
\usepackage{dcolumn} % Align table columns on decimal point
\usepackage{bm}
\usepackage{color}
\usepackage{epsfig}
\newcommand{\scfo}{Sr$_2$CrFeO$_6$}
\newcommand{\lcfo}{La$_2$CrFeO$_6$}
\begin{document}
\title{Evaluation of Half-metallic Antiferromagnetism in ${\cal A}_2$CrFeO$_6$ ({$\cal A$}=La, Sr)
}
\author{Kwan-Woo Lee$^{1,2}$} 
\email{mckwan@korea.ac.kr}
\author{Kyo-Hoon Ahn$^2$}
\affiliation{ 
$^1$Department of Display and Semiconductor Physics, 
  Korea University, Jochiwon, Chungnam 339-700, Korea \\
$^2$Department of Applied Physics, Graduate School, 
 Korea University, Jochiwon, Chungnam 339-700, Korea
}
\date{\today}
\begin{abstract}
The nearly well-ordered double perovskite La$_2$CrFeO$_6$ has been synthesized recently.
Contrary to previous theoretical predictions, but in agreement with 
experimental observations, our first principle calculations indicate
an insulating ferrimagnet La$_2$CrFeO$_6$ with antialigned $S$=$\frac{3}{2}$ Cr$^{3+}$ 
and $S$=$\frac{5}{2}$ Fe$^{3+}$ ions,
using the local spin density approximation (LSDA), a correlated band theory LDA+U, 
and a semilocal functional modified Becke-Johnson method.
Additionally, we investigated the double perovskite Sr$_2$CrFeO$_6$, 
which is as yet unsynthesized. In LSDA calculations, 
this system shows formally tetravalent Cr and Fe ions both having antialigned 
$S$=1 moments, but is a simple metal.
Once applying on-site Coulomb repulsion $U$ on both Cr and Fe ions, 
this system becomes half-metallic and 
the moment of Fe is substantially reduced, resulting in zero net moment. 
These results are consistent with our fixed spin moment studies. 
Our results suggest a precisely compensated half-metallic Sr$_2$CrFeO$_6$. 
\end{abstract}
\maketitle

\section{Introduction}
A half-metallic antiferromagnet has one conducting spin channel
and the other insulating spin channel, but shows zero net moment 
in a unit cell.\cite{groot} 
Contrary to the conventional antiferromagnet, 
antialigned moments of two different kinds of magnetic ions
are exactly compensated by each other.
Thus this is more properly called a compensated half-metal (CHM).\cite{felser}
Since CHM has no net moment, CHM is of extreme interest to spintronics and 
is anticipated to generate a single spin superconductor,\cite{wep1}
though no true CHM has yet been established.

After Pickett's prediction,\cite{wep2} 
double perovskite (DP) compounds have been investigated
as the most promising candidates for CHM. 
The DP of ${\cal A}_2$BB$^\prime$O$_6$ (${\cal A}$= La or Sr) has a rock-salt structure of 
transition metals BB$^\prime$.\cite{poeppel} 
In spite of many predictions of CHM 
in DP-related compounds,\cite{wep2,park,min,lee1,pp1,song,chen,pp2,lee2,lee3} 
no experimental evidence has been observed.
The reasons for this are not yet clear, but some seem to result from antisite disorder 
on B and B$^\prime$ sites.\cite{jana}
In particular, in systems having $3d$ ions in both B and B$^\prime$ sites, 
similar ionic radii often prevent a well-ordered sample from being established, 
resulting in undermined compensation of magnetic moments.

Ueda {\it et al.} synthesized artificial LaCrO$_3$/LaFeO$_3$ superlattices
along the [111] direction, proposing a ferromagnetic order
with $d^3$ Cr$^{3+}$ and $d^5$ Fe$^{3+}$.\cite{ueda}
Through measurements of the x-ray magnetic circular dichroism
in the superlattice sample, Gray {\it et al.}, who confirmed this trivalent 
oxidation state, obtained a smaller magnetic moment than a simple estimation 
of the spin-only magnetic moment and suggested a canted antiferromagnetic 
order.\cite{gray}
Very recently, Chakraverty {\it et al.} synthesized epitaxial 
La$_2$CrFeO$_6$ films using pulsed laser deposition, obtaining a nearly well-ordered
sample with at most 90 \% order in the B-sites.\cite{chak}
This sample shows the saturated moment of $\sim$2 $\mu_B$,
implying the antialigned trivalent Cr and Fe state.
The moment is consistent with Pickett's theoretical prediction
in the well-ordered phase,\cite{wep1} though the sample shows a small amount of
antisite disorder. This implies that the effects of such a small amount of 
disorder on the net moment are negligible in La$_2$CrFeO$_6$.

From the theoretical point of view, Pickett's calculations 
within the local spin density approximation (LSDA) indicated 
the half-metallic ferrimagnet (FI) of \lcfo.\cite{wep1} 
Miura and Terakura revisited
this compound using the generalized gradient approximation (GGA) and LDA+U.\cite{miura}
In their GGA calculations, this system had the same moments as Pickett's LSDA,
but was insulating.
Through LDA+U calculations, they concluded that \lcfo~ was insulating and ferromagnetic (FM).
However, the experimental observations show that this system 
is an insulating FI.\cite{chak,ohtomo}

In this paper, to disentangle this controversy, first we will revisit \lcfo~
using LSDA, LDA+U, the modified Becke-Johnson functional,\cite{mbj} 
and fixed spin moment (FSM) approach.\cite{fsm} 
Our results confirm the insulating FI \lcfo, consistent with the experimental
observations.\cite{chak}
Second, we will address the electronic and magnetic properties of 
the as-yet-unsynthesized DP \scfo, mainly using LDA+U which shows 
good agreement with the results of the experiment on \lcfo. 
One of the current authors and Pickett 
suggested a half semimetallic antiferromagnet in the isovalent 
and isostructural Sr$_2$Cr$\cal{T}$O$_6$ ($\cal{T}$=Os, Ru).\cite{lee1}
In the previous calculations with LSDA,\cite{lee1} 
\scfo~ was suggested to be a simple metallic ferrimagnet.
However, in our calculations including the Coulomb correlations, DP \scfo~ 
is a precisely compensated half-metal, as will be confirmed by FSM studies.

\section{Structure and calculation}
In the cubic DP (space group: $Fm\bar{3}m$, No. 225), 
the Fe and Cr ions sit on the $4a$ (0,0,0) 
and $4b$ ($\frac{1}{2}$,$\frac{1}{2}$,$\frac{1}{2}$) sites, respectively.
For \lcfo, we used the lattice parameter $a$=7.84 \AA~, which was 
estimated from the experiment.\cite{chak}
Within LSDA, the internal parameter of O ions, lying on the $24e$ sites ($x$,0,0),
is optimized. Our optimized internal parameter $x$=0.2508 implies
nearly identical oxidization states between the Fe and Cr ions.

For \scfo, our preliminary investigations using the Spuds program
indicate that the cubic DP is a stable structure.\cite{spuds}
Besides, considering tetravalent Cr and Fe ions (see below), 
the tolerance factor is
\begin{eqnarray}
t=\frac{r_A+r_O}{\sqrt2 (\frac{r_B+r_{B'}}{2}+r_O)}\approx1.02,\nonumber
\end{eqnarray}
unity for the ideal DP, using the Shannon ionic radius $r$.\cite{webelement} 
Thus, we fully optimized the lattice and internal parameters
of the cubic DP in this initial study. 
Using LSDA, our optimized lattice constant is $a$=7.447 \AA, about 15\% smaller
volume than \lcfo~ and Sr$_2$CrOsO$_6$.\cite{chak,krock}
The internal parameter $x$=0.2513 leads to a Cr-O bond length 
which is only 0.01 \AA~ shorter than the Fe-O bond length.
Here, we neglect such a small distortion as is observed in Sr$_2$CrOsO$_6$,\cite{krock}
since no significant change in the net moment results from this distortion.\cite{lee1}

Our calculations were carried out with LSDA and LDA+U approaches 
implemented in two all-electron full-potential
codes FPLO and WIEN2k. \cite{fplo1,wien2k}
In these LDA+U calculations, two popular double-counting schemes,
the so-called around mean field and fully localized limit,\cite{ldau1,ldau2} 
show similar results for these compounds.
The proper values of the on-site Coulomb repulsion $U$ in these systems
are unclear, but 3--4 eV for Cr ions\cite{scro} 
and 6 eV for Fe ions have been often used.\cite{shein}
(The Hund's exchange integral $J$=1 eV was fixed for all values of $U$,
since results obtained from $U_{eff}$=$U-J$ were identical with ones from
separate $U$ and $J$ inputs.)
FSM calculations implemented in FPLO 
were also carried out to investigate the robustness of our results.
In addition, the Becke-Johnson functional modified by Tran and Blaha
(the so-called mBJ),\cite{mbj} implemented in WIEN2k,
was used for comparison with our LDA+U results for \lcfo.
The semilocal functional mBJ is much cheaper than a hybrid 
functional, but provides reasonable agreement with experiments 
for transition metal oxides.\cite{mbj,singh_mbj}

The Brillouin zone was sampled with a regular mesh containing
up to 256 irreducible k-points.
The structural parameters were optimized until forces were smaller than
1 meV/\AA~ using FPLO.
In WIEN2k, the basis size was determined by $R_{mt}K_{max}$=7 and
the APW sphere radii: 2.5 for La, 1.95 for Cr, 1.96 for Fe, and 1.72 for O in \lcfo;
2.47 for Sr, 1.84 for Cr, 1.89 for Fe, and 1.64 for O in \scfo.

\begin{figure}[tbp]
%\vskip 8mm
{\resizebox{8cm}{6cm}{\includegraphics{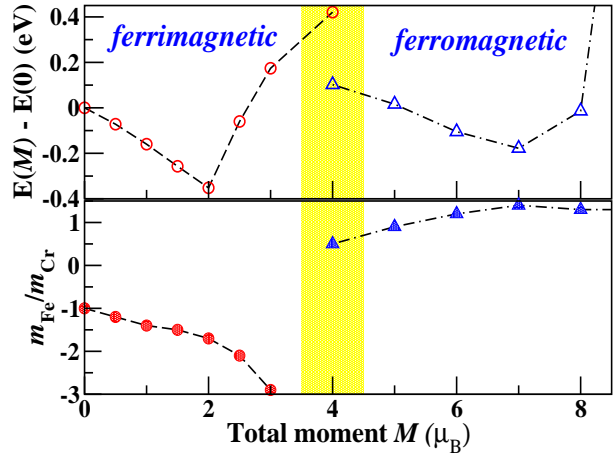}}}
\caption{(Color online) LSDA fixed spin moment calculations
in \lcfo. Top: energy vs. total moment $M$ plot.
The zero energy $E(0)$ denotes the energy of the exactly
compensated state.
Bottom: the ratio of local moments $m$ of Fe to Cr.
Below $M$$\sim$3.5 $\mu_B$ only FI states can be obtained,
while only FM states appear above $M$$\sim$4.5 $\mu_B$.
In the shaded regime, both FI and FM states exist.
The ground state at $M$=2 $\mu_B$ is an insulating FI state,
while the others are metallic.
}
\label{lafsm}
\end{figure}

\begin{figure}[tbp]
%\vskip 8mm
{\resizebox{8cm}{6cm}{\includegraphics{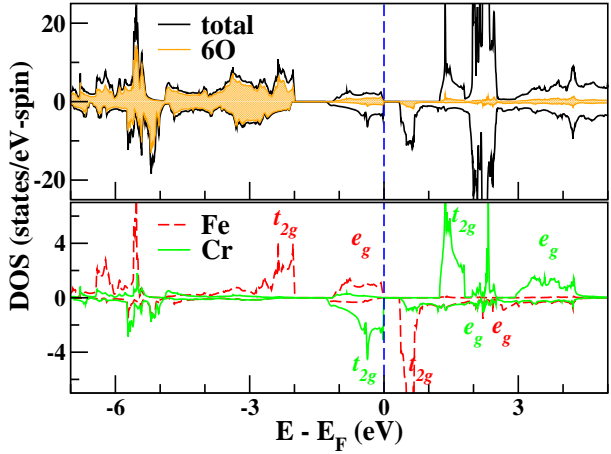}}}
\caption{(Color online) LSDA total and atom-projected densities of states 
(DOSs) in \lcfo, indicating an insulating FI with gaps of 1.2 and 0.3 eV
in the spin up and down channels. The sharp peaks around 2 eV correspond
to La $4f$ bands.
Bottom: $3d$-orbital projected DOSs in Fe and Cr ions.
The vertical dashed line denotes the Fermi energy $E_F$, which is set to zero.
}
\label{lados}
\end{figure}

\section{Revisited L\lowercase{a$_2$}C\lowercase{r}F\lowercase{e}O$_6$}
\subsection{LSDA ground state}
To unravel the discrepancies stated in the introduction,
we revisit \lcfo~ using various approaches: LSDA, LDA+U, mBJ, and FSM.
All of these calculations give consistent results (see below).

First, we will the address the FSM results within LSDA. 
The energy vs. total moment $M$ plot is given in Fig. \ref{lafsm}.
We obtained two solutions: FI below $M$$\approx$3.5 $\mu_B$ and
FM above $M$$\approx$4.5 $\mu_B$, while both states coexist in the region of
$M\approx$3.5--4.5 $\mu_B$. As $M$ is increased from 
the precisely compensated state ($M$=0), the energy decreases monotonically 
and reaches a minimum at $M$=2 $\mu_B$.
Below the minimum, the local moment of Cr, $m_{Cr}$$\approx$--2.4 $\mu_B$, 
is nearly unchanged, while that of Fe increases monotonically.
Above the minimum, the magnitude of $m_{Cr}$ decreases and finally becomes
almost zero at $M$=4.
The minimum state is insulating, leading to 
nonanalytic behavior in the FSM curve near the minimum as observed in
half-metals.\cite{lee1,lee_fsm} 
In the FM regime, the minimum energy occurs at $M$$\sim$ 7.
As $M$ is reduced from the minimum, the ratio $m_{Fe}/m_{Cr}$ 
of the local moments linearly decreases.
This FM minimum state has higher energy by 175 meV than 
the FI minimum state.
Thus, our FSM calculations indicate that the ground state is the insulating
FI state of $M$=2 $\mu_B$, in agreement with our self-consistent 
calculations (see below).

In LSDA calculations, we obtained both FM and FI states.
As mentioned previously, FI is energitically favored over FM.
This FM state has the total moment of $M$=7.23 $\mu_B$ (3.90 for Fe,
2.74 for Cr, and 0.51 for the six oxygens).
In the FI state, the total moment of $M$=2 is decomposed 
into local moments of 3.94 for Fe, --2.37 for Cr, and 0.24 for the six oxygens
(in units of $\mu_B$).
Thus this system has nominally high spin Fe$^{3+}$ 
($t_{2g}^{3\uparrow}$$e_{g}^{2\uparrow}$) and Cr$^{3+}$ ($t_{2g}^{3\downarrow}$),
as is also visible in DOSs given in Fig. \ref{lados}.
Note that the strong $p-d$ hybridization, observable around --6 eV,
substantially reduces the Fe moment.
The $t_{2g}$-$e_g$ crystal field splitting of Fe is 2 eV, about 1 eV 
smaller than the exchange splitting, leading to high spin $S$=$\frac{5}{2}$ 
Fe$^{3+}$. Both the $t_{2g}$-$e_g$ crystal field and exchange splittings of 
Cr are about 2 eV.
Compared with the metallic FM state,
the antiferromagnetic interactions between Fe and Cr ions lead to gaps
in both spin channels, as shown in Fig. \ref{lados}. 
So this system becomes a band insulator, consistent with the experiment.\cite{chak}
Note that the nonmagnetic state has a much higher energy than these two states.

\begin{figure}[tbp]
%\vskip 8mm
{\resizebox{8cm}{6cm}{\includegraphics{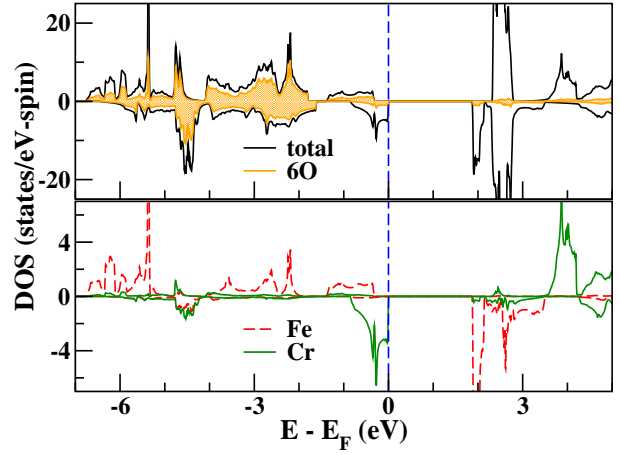}}}
\caption{(Color online) mBJ total and atom-projected DOSs
in FI \lcfo, showing a gap of 1.8 eV in the spin down channel.
The unoccupied Cr and Fe $t_{2g}$ manifolds in the spin up and down
channels are at 4 eV and 2 eV, respectively.
}
\label{mbjdos}
\end{figure}

\subsection{Inclusion of correlation effects}
Miura and Terakura claimed 
that the ground state was an insulating FM through LDA+U calculations
using a pseudopotential code,\cite{miura}
in contrast to the experimental observations.
We investigated this system using the LDA+U approach, implemented 
in two all-electron full-potential codes FPLO and WIEN2k.\cite{fplo1,wien2k}
We considered the Coulomb correlation of both transition metals
in the range of $U$=3--7 eV.
As shown in the calculations of Miura and Terakura,
two insulating FM and FI states were obtained.
However, in our calculations, the FI with $M$=2 $\mu_B$ 
is energetically favored  over the FM state with $M$=8 $\mu_B$, 
regardless of the strength of the on-site Coulomb repulsion $U$.
In particular, at $U$=4 eV for Cr and 6 eV for Fe, 
the difference in energy between the states is 86 meV.

To confirm our results, we carried out mBJ calculations,
which give the correct magnetic ground state in a parent compound of cuprate
CaCuO$_2$.\cite{singh_mbj}
Consistently with our other calculations described above,
the FI state has a lower energy by 180 meV than the FM state, 
indicating that our results are very robust and show good agreement
with the experiment.\cite{chak}
The mBJ total and atom-projected DOSs are given in Fig. \ref{mbjdos}.
Compared with LSDA results shown in Fig. \ref{lados}, 
the unoccupied $t_{2g}$ manifolds of Cr in the spin up and Fe
in the spin down considerably shift up,
while the occupied bands show very similar features.
The local moments are 4.11 for Fe, --2.42 for Cr, and
0.60 for the six oxygens, again resulting in the net moment of 2 
(in units of $\mu_B$).
The gaps are 2.5 eV and 1.8 eV for the spin up and down channels,
respectively.

\begin{figure}[tbp]
%\vskip 8mm
{\resizebox{8cm}{6cm}{\includegraphics{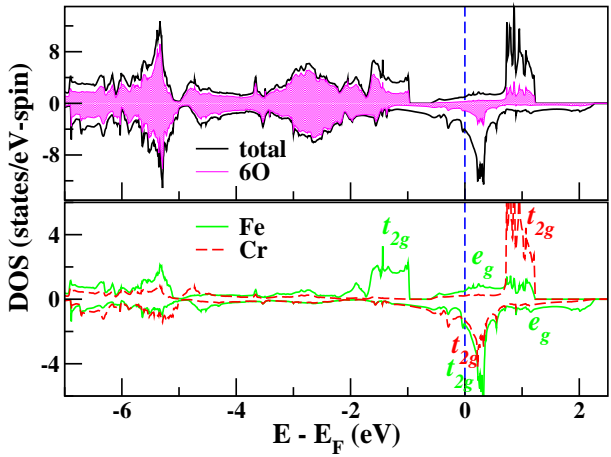}}}
\caption{(Color online) LSDA total and atom-projected DOSs in \scfo, 
 indicating a FI state. 
 The O states lie in the range of --8 to --1.8 eV.
 The unfilled Cr $e_g$ manifolds in both spin channels exist above 3 eV 
 (not shown here).
 The density of states $N(E_F)$ at $E_F$ is 5.08 states per eV.
}
\label{ldados}
\end{figure}

\section{Electron structure of S\lowercase{r$_2$}C\lowercase{r}F\lowercase{e}O$_6$}
\subsection{LSDA electronic structure}
The LSDA total and atom-projected DOSs are represented in Fig. \ref{ldados}.
In the spin down channel, the partially filled $t_{2g}$ manifolds
of both Cr and Fe lie in the region of --1 to 1 eV, 
and come into contact with the unfilled Fe $e_{g}$ manifold.
This mixing between Cr and Fe $t_{2g}$ manifolds remains significant
even for inclusion of correlation effects (see below), 
indicating an unusual charge transfer.
In the up channel, the completely filled Fe $t_{2g}$ manifold 
lies in the range of --1.7 to --1 eV, while the center of 
the unfilled Cr $t_{2g}$ manifold with the width of 0.5 eV 
is at 1 eV. 
For Fe ions, the $t_{2g}$-$e_g$ crystal field splitting of 1.8 eV in the spin up 
is accidentally equal to the exchange splitting of the $t_{2g}$ manifold.
So, some of the Fe $e_{g}$ bands cross over $E_F$, undermining the half-metallicity.
In this state, the Cr moment of --1.32 $\mu_B$ is antialigned 
with those of Fe and the oxygens (1.86 for Fe and 0.21 for 6O), 
resulting in the total moment 0.75 $\mu_B$.
These results indicate formally tetravalent $d^2$ Cr and $d^4$ Fe ions
of $S$=1 both in this system, 
in contrast to trivalent Cr and pentavalent $d^3$ Os/Ru in Sr$_2$Cr$\cal{T}$O$_6$.
Within LSDA these distinctions between the systems lead to metallic \scfo, 
but half-semimetallic ({\it i.e.}, band insulating) Sr$_2$Cr$\cal{T}$O$_6$.
Thus, this requires the inclusion of correlations to \scfo, as is usual in 3d systems.
In the next subsection, we will address this issue.

This FI state is energetically favored over the nonmagnetic
state by 283 meV, consistent with our FSM calculations (see below).
Attempts to obtain a FM state always converged to the
FI or nonmagnetic state.
Thus, the state of the antialigned moments is very stable.

\begin{figure}[tbp]
%\vskip 8mm
{\resizebox{8cm}{6cm}{\includegraphics{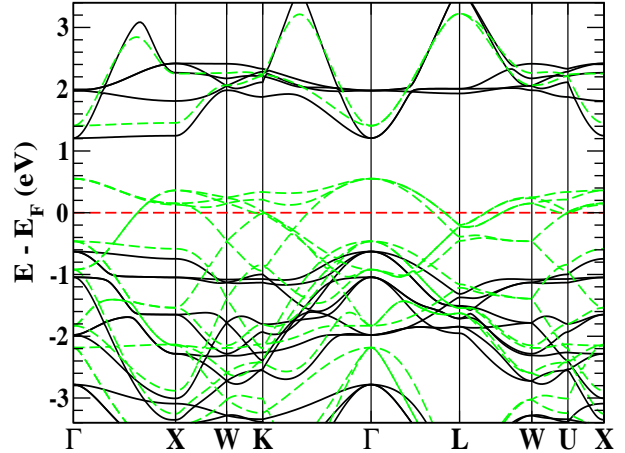}}}
\caption{(Color online) LDA+U band structure for both the spin up (solid)
 and down (dashed) channels in \scfo, at $U$= 4 eV for Cr and 6 eV for Fe,
 showing a half-metal.
 The horizontal dashed line denotes $E_F$, which is set to zero.
}
\label{uband}
\end{figure}

\begin{figure}[tbp]
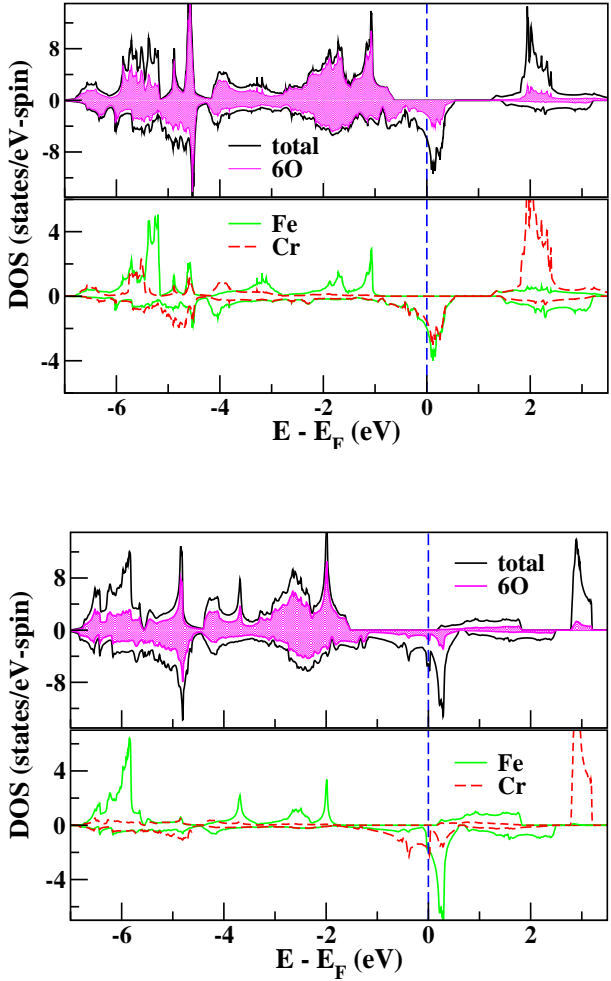

%\vskip 8mm
{\resizebox{8cm}{6cm}{\includegraphics{Fig6a.eps}}}
\vskip 10mm
{\resizebox{8cm}{6cm}{\includegraphics{Fig6b.eps}}}
\caption{(Color online) \scfo~ total and atom-projected DOSs 
 in (top) LDA+U and in (bottom) mBJ.
 In both results, a gap of $\sim$1.8 eV appears in the spin up.
 In the spin down, the partially filled $t_{2g}$ manifolds of
 both Fe and Cr ions cross over $E_F$. 
 In the spin up, the Fe $t_{2g}$ manifold is fully filled,
 while The Cr $t_{2g}$ manifold is completely unfilled.
 This indicates formally Cr$^{4+}$ $t_{2g}^{2\downarrow}$
 and Fe$^{4+}$ $t_{2g}^{3\uparrow, 1\downarrow}$,
 though the concept of formal charge is some murky
 due to strong hybridization.
}
\label{udos}
\end{figure}

\subsection{Correlated electronic structure}
In LDA+U calculations, as in \lcfo, we used values of $U$
in the range of 3--7 eV for both Cr and Fe ions, showing consistent results.
Now, we will focus on results obtained at $U$= 4 eV for Cr and 6 eV for Fe,
which have been widely used in perovskite-type oxides.\cite{scro,shein}

Figure \ref{uband} shows an enlarged band structure near $E_F$,
indicating a half-metal.
The mixture of Cr $t_{2g}$ and $e_{g}$ manifolds in the spin up and the Fe $e_g$
manifold in the spin down lies in the range from 1 to 3.5 eV.
In particular, the Cr $t_{2g}$ around 2 eV is nearly dispersionless.
In the spin up (insulating) channel, the top of the valence bands has mostly 
O $p$ character, 
leading to a gap of 1.8 eV between the O $p$ and the Cr $e_g$ bands,
as clearly shown in the total and atom-projected DOSs of the top panel of
Fig. \ref{udos}.
The Fe $t_{2g}$ manifold in the spin up is completely filled.
In the spin down (conducting) channel, the overlapped $t_{2g}$ manifolds 
of Fe and Cr are partially filled. In this regime, each character of Cr and Fe is
nearly identical to that of the oxygens, indicating strong $p-d$ hybridization.
This substantial hybridization is also visible in DOSs of the top panel 
of Fig. \ref{udos}.
Applying $U$ to both Fe and Cr ions leads to a charge transfer of 0.5$e$
(which is measured from the Mullikan charge decomposition in the FPLO method) 
from the six oxygens to the Fe ion, subsequently enhancing the $p$-$d$ hybridization.
As a result, the moment of Fe remarkably reduces to 0.98 $\mu_B$, but
that of the six oxygens increases to 0.42 $\mu_B$.
However, the Cr moment remains nearly unchanged, resulting in zero net moment.

We also carried out mBJ calculations in this system.
The bottom panel of Fig. \ref{udos} shows the corresponding DOSs,
reproducing the half-metallicity.
Compared with LDA+U results, the unfilled Fe $e_{g}$ manifolds shift toward $E_F$
in both spin channels, leading to decreasing the $t_{2g}$-$e_{g}$ 
crystal field splitting of Fe ions.
In the spin down, the partially filled Fe $t_{2g}$ manifold moves upward,
while the Cr $t_{2g}$ manifold shifts down.
So, the unusual mixing between these $t_{2g}$ manifolds, which is observed
in LSDA and LDA+U, is considerably diminished, though some amount of the mixing 
is still visible.
However, the net moment remains zero: 1.80 for Fe, --1.96 for Cr, and
0.36 for 6O (in units of $\mu_B$).
Thus, this system is a precisely compensated half-metal.

\begin{figure}[tbp]
%\vskip 8mm
{\resizebox{8cm}{6cm}{\includegraphics{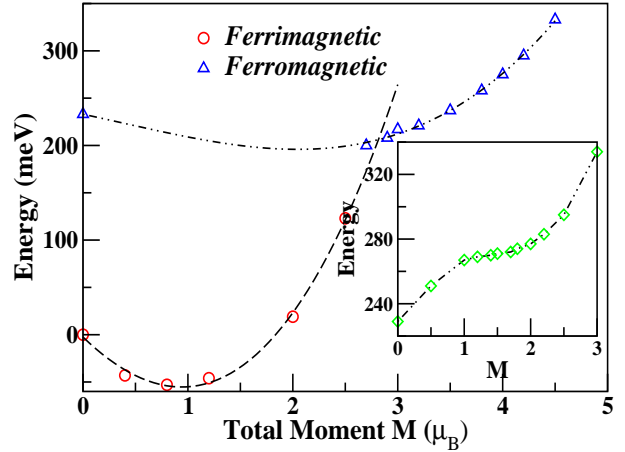}}}
\caption{(Color online) Fixed spin moment (FSM) calculations of LSDA in \scfo.
 The symbols of circles and triangles indicate FI
 and FM states, respectively.
 The zero energy is set to the energy of an exactly compensated 
 FI state with $M$=0. 
 {\it Inset}: FSM of LDA+U at $U$=4 and 6 eV for Cr and Fe ions, respectively.
 This shows a change in energy for FM, except for nonmagnetic state at $M$=0. 
 A plateau appears in the regime of $M$$\approx$1 to 2 $\mu_B$.
 The zero energy denotes CHM, which is the ground state 
 in the LDA+U calculations. 
}
\label{fsm}
\end{figure}

\subsection{Fixed spin moment studies}
To investigate the robustness of this compensated half-metal,
we carried out FSM calculations.\cite{fsm}
The change in energy vs. total moment $M$ plot is given in Fig. \ref{fsm}.
In LSDA, we obtained two curves.
In the range of $M$=0 to 2.5 $\mu_B$ FI solutions are obtained, 
while FM solutions are obtained above 2.7 $\mu_B$.
The FI state with $M$=0.75 $\mu_B$ is
the ground state, about 50 meV lower in energy than CHM. 
At $M$=0, the nonmagnetic state has 230 meV higher energy than CHM.
So, a state having antialigned moments of Cr and Fe ions is energetically favored.

The inset of Fig. \ref{fsm} shows FSM results for FM states in LDA+U.
Note that the CHM state with $M$=0, which is set to the zero energy, 
has lowest energy.
Contrary to the LSDA results, FI states could be obtained only 
near $M$$\approx$0 (not shown here).
Interestingly, a plateau appears in the range of $M$=1 to 2 $\mu_B$.
Our self-consistent calculations also produce a FM state having $M$$\approx$2
in LDA+U.
This implies that a meta-stable state may appear in the regime of a plateau.

\begin{figure}[tbp]
%\vskip 8mm
\rotatebox{-90}{{\resizebox{6cm}{8cm}{\includegraphics{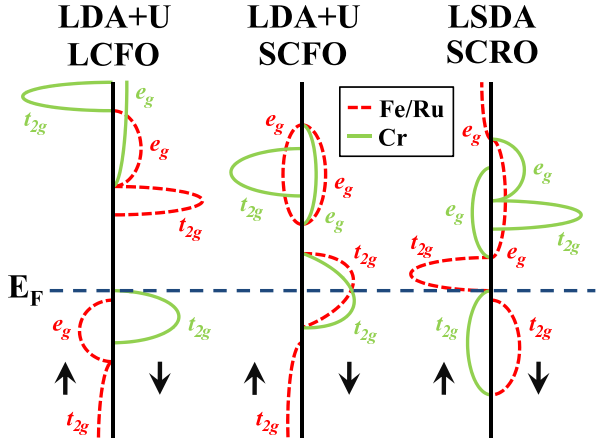}}}}
\caption{(Color online) Simplified Cr and Fe atom-projected DOSs of \lcfo~ 
and \scfo~ of LDA+U, compared with Cr- and Ru-projected DOSs of LSDA 
Sr$_2$CrRuO$_6$ which has been predicted to be a compensated half-semimetal.\cite{lee2}
In these systems, Cr ions follow the first Hund's rule.
}
\label{schemdos}
\end{figure}

\section{Discussion and Summary}
Figure \ref{schemdos} displays the schematic atom-projected DOSs of both transition metals
in \lcfo, \scfo, and Sr$_2$CrRuO$_6$, indicating two interesting aspects.
First, Fe ions are in the $d^5$ $S$$=\frac{5}{2}$ high spin state in \lcfo, 
but in the $d^4$ $S$=1 low spin state in \scfo.
To uncover the origin of the transition of the Fe spin states  
when replacing La with Sr, we carried out calculations of \scfo~ with the same structure as \lcfo.
Our results show that \scfo~ has a similar electronic structure to \lcfo,
with the only difference being the band filling,
resulting in a simple ferrimagnetic metal with a $d^4$ $S$=2 high spin Fe ion. 
This indicates that the 15\% volume in \scfo~ reduction leads to an increase in
the crystal field splitting, and subsequently the transition.
Second, \scfo~ is half-metallic, while an isovalent and isostructural Sr$_2$CrRuO$_6$ 
is semimetallic. 
CHM \scfo~ remains unchanged even for our calculations with the same structure as Sr$_2$CrRuO$_6$.
Thus, this distinction results from chemical differences between Fe and Ru ions rather than
differences in the crystal structure.

In summary, we revisited a well-ordered DP \lcfo, which shows noticeable discrepancies
between the predictions of existing models and the experimental results, 
through various first principles approaches implemented in two all-electron
full-potential codes.
Our calculations show good agreements with the experiment,\cite{chak} indicating
an insulating ferrimagnet with $M$=2 $\mu_B$.
Furthermore, we investigated the unsynthesized DP \scfo. 
Our LSDA+U results suggest a compensated half-metallic \scfo~
with nominal Cr$^{4+}$ ($t_{2g}^{2\downarrow}$)
and Fe$^{4+}$ ($t_{2g}^{3\uparrow, 1\downarrow}$), which is confirmed 
by the FSM calculations.
This CHM \scfo~ is expected to be synthesized by the recently developed 
method that is used to synthesize \lcfo~ and Sr$_2$CrOsO$_6$.\cite{chak,krock}

\section{Acknowledgements}
We acknowledge useful communication with A. Ohtomo on the experimental observations. 
This research was supported by the Basic Science Research Program through
NRF of Korea under Grant No. 2012-0002245.

\end{document}